\documentstyle[aps,12pt]{revtex}

\begin{document}
\title{Domain Bubbles of Extra Dimensions}
\author{J.R. Morris\thanks{%
E-mail: jmorris@iun.edu}}
\address{{\it Physics Dept., Indiana University Northwest,}\\
{\it 3400 Broadway, Gary, Indiana 46408}\\
\smallskip\ }
\maketitle

\begin{abstract}
``Dimension bubbles'' of the type previously studied by Blau and Guendelman
[S.K. Blau and E.I. Guendelman, Phys. Rev. D40, 1909 (1989)], which
effectively enclose a region of 5d spacetime and are surrounded by a region
of 4d spacetime, can arise in a 5d theory with a compact extra dimension
that is dimensionally reduced to give an effective 4d theory. These bubbles
with thin domain walls can be stabilized against total collapse in a rather
natural way by a scalar field which, as in the case with ``ordinary''
nontopological solitons, traps light scalar particles inside the bubble.

\smallskip\ 


\bigskip\ 

\newpage\ 
\end{abstract}


\section{Introduction}

Blau and Guendelmann\cite{BG} have investigated the interesting case of an
inhomogeneous spacetime formed by sewing together regions having different
numbers of macroscopic spatial dimensions. From a four dimensional (4d)
point of view the scale factor of a fifth (compact) dimension can be treated
as a scalar field which is allowed to vary rapidly in a localized region of
the 4d spacetime, forming a domain wall. The domain wall can thereby
interpolate between two regions differing in the number of macroscopic space
dimensions. Closed domain walls can form ``dimension bubbles'' with
different numbers of macroscopic space dimensions on the interior and
exterior of the bubble wall. As examples, Blau and Guendelmann\cite{BG} have
used the low temperature Rubin-Roth 4d effective potential, generated by
including one-loop corrections due to quantum fluctuations from fermions and
bosons\cite{RR}. Fermions, which can have Casimir energies of opposite sign
from bosonic ones, can act toward stabilizing a compact extra dimension\cite
{RR} from collapsing due to the gravitational Casimir effect\cite{AC}. From
a 4d point of view, the domain wall arises from the scalar dilaton field $%
\varphi $ associated with the scale factor of the extra dimension, and the
Rubin-Roth potential becomes a function of this field.

There is a range of parameters for which the 4d effective potential $%
U(\varphi )$ exhibits one local minimum, at say, $\varphi =\varphi _{\min }$%
, separated by a potential barrier from another minimum approached
asymptotically as $\varphi \rightarrow \infty $. This type of potential
gives rise to what will be referred to here as a ``{\it semi-vacuumless}''
domain wall, since it seems to be a hybrid between an ordinary domain wall
and a ``vacuumless'' domain wall of the type originally described by
Vilenkin and Cho\cite{CV} and further studied by Bazeia\cite{Baz}. The
scalar field $\varphi $ interpolates between the vacuum domain ($\varphi
=\varphi _{\min }$) and the ``vacuumless'' domain ($\varphi \rightarrow
\infty $). If the values of the effective potential $U(\varphi )$ differ at
all on the two different sides of the wall, the wall will generally be
unstable against bending. As a result, a network of closed domain bubbles
can form. Such a bubble, however, will be unstable against collapse without
a mechanism to balance the inward pressure due to the tension in the bubble
wall.

A simple mechanism proposed here to stabilize one of these
``semi-vacuumless'' bubbles consists of the inclusion of a fundamental
scalar field $\chi $ in the original 5d action. Upon dimensional reduction
to the effective 4d (Einstein frame) theory, this field $\chi $ is seen to
have a mass which depends upon $\varphi $, so that lower mass $\chi $ bosons
in the interior of the bubble can be effectively trapped, as is the case
with an ``ordinary'' nontopological soliton (NTS), as previously studied by
Frieman, Gleiser, Gelmini and Kolb\cite{NTS}. The trapped $\chi $ particles
give rise to an outward pressure that can stabilize the bubble from total
collapse. We specifically consider bubbles, in the thin wall approximation,
with $\varphi =\varphi _{\min }$ outside the bubble and $\varphi =\varphi
_1\gg \varphi _{\min }$ inside the bubble. The scale factor $B$ of the extra
dimension is related to the scalar $\varphi $ by $B=e^{\sqrt{\frac 23}\kappa
\varphi }$, with $\kappa =\sqrt{8\pi G}$, so that for $\varphi _1\rightarrow
\infty $ the extra dimension becomes macroscopic in the bubble's interior,
even if the bubble has a small size in the surrounding 3d space. This object
is then a ``dimension bubble'' embedded in an effective 4d spacetime with an
effective 5d interior (for a macroscopic bubble). (A microscopic bubble
would have only one macroscopically large space dimension inside.) As with
ordinary NTSs, these bubbles, if physically realized, could contribute to
the dark matter content of our universe.

The rest of the paper is organized in the following way: The dimensional
reduction of the 5d action is presented in sec. 2, allowing an extraction of
the basic form of the 4d effective potential. The 4d effective potential is
written in terms of the Rubin-Roth potential and a 5d cosmological constant.
The contributions to the low temperature Rubin-Roth potential are listed,
which describe contributions from quantum fluctuations for bosons and
fermions. Semi-vacuumless domain walls arising from the 4d effective
potential along with the instability of dimension bubbles are considered in
sec. 3. The stabilization mechanism is presented in sec. 4, describing $\chi 
$-boson stabilized NTS bags. Some properties of these NTS bags, such as the
bag mass and limits on the bag charge and radius, are investigated. A brief
summary forms sec. 5.

\section{Dimensional Reduction of the Five-Dimensional Action}

\subsection{Metric Ansatz}

We begin by considering a five-dimensional spacetime described by a metric $%
\tilde{g}_{MN}$: 
\begin{equation}
ds^2=\tilde{g}_{MN}dx^Mdx^N=\tilde{g}_{\mu \nu }dx^\mu dx^\nu +\tilde{g}%
_{55}dy^2  \label{e1}
\end{equation}

\noindent where $x^M=(x^\mu ,y)$, with $M,N=0,\cdot \cdot \cdot 3,5$ and $%
\mu ,\nu =0,\cdot \cdot \cdot 3$. We denote $\tilde{g}=\det \tilde{g}_{\mu
\nu }$, and $\tilde{g}_5=\det \tilde{g}_{MN}$, so that $\,\sqrt{|\tilde{g}_5|%
}=\sqrt{-\tilde{g}}\sqrt{|\tilde{g}_{55}|}$. The extra dimension will be
assumed to be toroidally compact, so that the 5d spacetime has topology of $%
M_4\times S^1$.

We assume an ansatz where the metric $\tilde{g}_{MN}$ is independent of the
extra dimension $y$, i.e., $g_{MN}=g_{MN}(x^\mu )$, $\partial _5g_{MN}=0$,
and the metric factorizes with $g_{\mu n}=0$. The extra dimension is
described by a linear coordinate $y$ lying in the range $0\leq y\leq 2\pi R$
and can be assigned a (dimensionless) scale factor $B(x^\mu )$ with $\tilde{g%
}_{55}=-B^2$. A scalar field $\varphi $ can be defined by 
\begin{equation}
\varphi =\frac 1\kappa \sqrt{\frac 32}\ln B,  \label{e2}
\end{equation}

\noindent where $\kappa $ is related to the 4d Planck mass $M_P$ by $\kappa =%
\sqrt{8\pi G}=\sqrt{8\pi }M_P^{-1}$, so that the scale factor can be written
as $B=e^{\sqrt{\frac 23}\kappa \varphi }$.

\subsection{The 5d Action and Dimensional Reduction}

We take the 5-dimensional action to include the 5d Einstein action,
cosmological constant $\Lambda $, and a source Lagrangian ${\cal L}_5$: 
\begin{equation}
\begin{array}{ll}
S_5 & =%
{\displaystyle {1 \over 2\kappa _5^2}}
\int d^5x\sqrt{\tilde{g}_5}\left\{ \tilde{R}_5-2\Lambda +2\kappa _5^2{\cal L}%
_5\right\} \\ 
& =%
{\displaystyle {V_y \over 2\kappa _5^2}}
\int d^4x\sqrt{-\tilde{g}}(B)\left\{ \tilde{R}_5-2\Lambda +2\kappa _5^2{\cal %
L}_5\right\} \\ 
& =%
{\displaystyle {1 \over 2\kappa ^2}}
\int d^4x\sqrt{-\tilde{g}}(B)\left\{ \tilde{R}_5-2\Lambda +2\kappa ^2{\cal L}%
\right\}
\end{array}
\label{e3}
\end{equation}

\noindent where we have used the definitions $V_y=\int dy=(2\pi R)$, $\kappa
_5^2=8\pi G_5=V_y\kappa ^2$, and ${\cal L}=V_y{\cal L}_5=(2\pi R){\cal L}_5$%
. In addition, $\tilde{R}_5=\tilde{g}^{MN}\tilde{R}_{MN}$ denotes the
5-dimensional Ricci scalar built from $\tilde{g}_{MN}$. Note that the 4d
Jordan Frame metric is $\tilde{g}_{\mu \nu }$, the $\mu \nu $ part of $%
\tilde{g}_{MN}$. A 4d Einstein Frame metric $g_{\mu \nu }$ can be defined by 
$g_{\mu \nu }=B\tilde{g}_{\mu \nu }=e^{\sqrt{\frac 23}\kappa \varphi }\tilde{%
g}_{\mu \nu }$, in which case the line element in (\ref{e1}) takes the
Kaluza-Klein form 
\begin{equation}
\begin{array}{ll}
ds^2 & =B^{-1}g_{\mu \nu }dx^\mu dx^\nu -B^2dy^2 \\ 
& =e^{-\sqrt{\frac 23}\kappa \varphi }g_{\mu \nu }dx^\mu dx^\nu -e^{2\sqrt{%
\frac 23}\kappa \varphi }dy^2
\end{array}
\label{e4}
\end{equation}

Using (\ref{e3}) and (\ref{e4}), the 5d action is dimensionally reduced to
the effective 4d Einstein Frame action 
\begin{equation}
S=\int d^4x\sqrt{-g}\left\{ \frac 1{2\kappa ^2}R+\frac 12(\nabla \varphi
)^2+e^{-\sqrt{\frac 23}\kappa \varphi }[{\cal L}-\frac 1{\kappa ^2}\Lambda
]\right\}  \label{e5}
\end{equation}

\noindent where $R=g^{\mu \nu }R_{\mu \nu }$ is the 4d Ricci scalar built
from the 4d Einstein Frame metric $g_{\mu \nu }$ and $g=\det g_{\mu \nu }$.
[We use a 4d metric with signature $(+,-,-,-)$.]

\subsection{The Form of the 4d Effective Potential}

In order to investigate the effects of Casimir-like one loop quantum
corrections on the size of the extra dimension, we follow Blau and Guendelman%
\cite{BG} and use the low temperature limit of the finite-temperature
Rubin-Roth potential $V_{RR}$ for bosons and fermions\cite{RR}. From this
potential, along with a cosmological constant term, a 4d effective potential 
$U$ can be constructed. The part of the original action containing the
potential terms, from (\ref{e3}), is 
\begin{equation}
S_{5,\text{ pot}}=-\int d^5x\sqrt{\tilde{g}_5}\left\{ V_5+\frac \Lambda {%
\kappa _5^2}\right\} =-\int d^4x\sqrt{-\tilde{g}_4}\int dy\left| \tilde{g}%
_{55}\right| ^{1/2}\left\{ V_5+\frac \Lambda {\kappa _5^2}\right\}
\label{e6}
\end{equation}

\noindent where $V_5$ is the potential appearing in ${\cal L}_5={\cal L}%
/(2\pi R)$. Using (\ref{e5}), we identify the 4d Einstein Frame effective
potential $U$ as 
\begin{equation}
U=\frac 1B\left[ V+\frac \Lambda {\kappa ^2}\right] =e^{-\sqrt{\frac 23}%
\kappa \varphi }\left[ V+\frac \Lambda {\kappa ^2}\right]  \label{e7}
\end{equation}

\noindent where $V=V_5(2\pi R)$ is the potential term appearing in ${\cal L}$%
. Identifying the Rubin-Roth potential through the relation 
\begin{equation}
V_{RR}\propto \int dy\left| \tilde{g}_{55}\right| ^{1/2}V_5=BV  \label{e8}
\end{equation}

\noindent so that $V\propto \frac{V_{RR}}B$, allows the 4d effective
potential to be written in terms of $V_{RR}$ as\footnote{%
The Rubin-Roth potential $V_{RR}$ is calculated in the Jordan Frame with a
Minkowski background in ref.\cite{RR}. A factor of $B^{-2}=\exp (-2\sqrt{%
\frac 23}\kappa \varphi )$ then accompanies $V_{RR}$ in the expression for
the Einstein Frame effective potential $U$. See ref.\cite{BG} for a basic
description of the effective potential and its dependence upon parameters.} 
\begin{equation}
U\propto \frac 1B\left[ \frac{V_{RR}}B+\frac \Lambda {\kappa ^2}\right] =e^{-%
\sqrt{\frac 23}\kappa \varphi }\left[ e^{-\sqrt{\frac 23}\kappa \varphi
}V_{RR}+\frac \Lambda {\kappa ^2}\right]  \label{e9}
\end{equation}

\subsection{Low Temperature Rubin-Roth Potential (per degree of freedom)}

Following the example of Blau and Guendelman, we consider a potential with
an interesting structure generated by Casimir contributions from the
graviton and additional fermionic matter fields\cite{BG}. We are interested
in the limit of negligible temperature, i.e., the high $\beta =1/T$ limit.
The asymptotic behaviors of bosonic and fermionic contributions to the
Rubin-Roth potential (per degree of freedom) are listed in Table 1 of ref.%
\cite{RR}. The results for massless bosons and massive fermions are listed
here for convenience. The potential terms can be written in terms of the
circumference of the extra dimension, $L_5=(2\pi R)B$.

{\bf Massless Bosons} \thinspace \thinspace \thinspace \thinspace \thinspace
For massless bosons, we have (for all $L_5$), 
\begin{equation}
V_{RR}^{(b)}\sim -\frac{3\zeta (5)}{4\pi ^2}\frac \beta {L_5^4},  \label{e10}
\end{equation}

\noindent where the scale factor $B$ is related to $L_5$ by 
\begin{equation}
\left( \frac{L_5}{2\pi R}\right) =|\tilde{g}_{55}|^{1/2}=B=e^{\sqrt{\frac 23}%
\kappa \varphi }  \label{e11}
\end{equation}

{\bf Fermions (}$M\ge 0${\bf )} \thinspace \thinspace \thinspace \thinspace
We can consider the low temperature (high $\beta $) limit with $L_5\ll \beta 
$ and $\beta \gg 1/M$ (i.e., $T\ll M$, for $M>0$) and the two cases of
(1)small $L_5$ ($L_5\ll 1/M$) and (2)large $L_5$ ($L_5\gg 1/M$):

\begin{equation}
V_{RR}^{(f)}\sim \left\{ 
\begin{array}{ll}
{\displaystyle {3\zeta (5) \over 4\pi ^2}}
{\displaystyle {\beta \over L_5^4}}
, & L_5\ll 
{\displaystyle {1 \over M}}
,\,\,M\geq 0 \\ 
{\displaystyle {M^2 \over 4\pi ^2}}
{\displaystyle {\beta \over L_5^2}}
e^{-ML_5}, & L_5\gg 
{\displaystyle {1 \over M}}
,\,\,M>0
\end{array}
\right\}  \label{e12}
\end{equation}

Assume now that there are $N_b$ massless bosonic degrees of freedom with $%
N_b\geq 5$, (5 graviton degrees of freedom) since, at least the graviton,
contributes and $N_f$ fermionic degrees of freedom, with $N_f>N_b$ (as in
ref.\cite{BG}). We then write the net Rubin-Roth potential as 
\begin{equation}
V_{RR}=N_bV_{RR}^{(b)}+N_fV_{RR}^{(f)}  \label{e13}
\end{equation}

\noindent so that, with the help of (\ref{e10}) and (\ref{e12}), we have 
\begin{equation}
V_{RR}\sim \left\{ 
\begin{array}{ll}
(N_f-N_b)%
{\displaystyle {3\zeta (5) \over 4\pi ^2}}
{\displaystyle {\beta \over L_5^4}}
, & L_5\ll 
{\displaystyle {1 \over M}}
\\ 
-N_b%
{\displaystyle {3\zeta (5) \over 4\pi ^2}}
{\displaystyle {\beta \over L_5^4}}
+N_f%
{\displaystyle {M^2 \over 4\pi ^2}}
{\displaystyle {\beta \over L_5^2}}
e^{-ML_5}, & L_5\gg 
{\displaystyle {1 \over M}}
\end{array}
\right\}  \label{e14}
\end{equation}

{\it Note}: For fermions with different masses, we can make the replacement 
\begin{equation}
N_f\frac{M^2}{4\pi ^2}\frac \beta {L_5^2}e^{-ML_5}\rightarrow 
\mathop{\displaystyle \sum }
_i\left\{ N_f^{(i)}\frac{M_i^2}{4\pi ^2}\frac \beta {L_5^2}%
e^{-M_iL_5}\right\}  \label{e15}
\end{equation}

\noindent with the index $i$ running over the different fermionic species.

\subsection{The Four Dimensional Effective Potential}

Using (\ref{e9}) and (\ref{e11}) the 4d Einstein Frame effective potential $%
U $ can be written in terms of the circumference of the extra dimension, $%
L_5 $, as 
\begin{equation}
U=c_1^{\prime }\frac{V_{RR}}{L_5^2}+\frac{c_2}{L_5}\frac \Lambda {\kappa ^2},
\label{e16}
\end{equation}

\noindent where $c_1^{\prime }$ and $c_2$ are positive constants. Using the
asymptotic forms of (\ref{e14}) the effective potential takes the form

\begin{equation}
U\sim \left\{ 
\begin{array}{ll}
c_1(N_f-N_b)%
{\displaystyle {3\zeta (5) \over 4\pi ^2}}
{\displaystyle {\beta \over L_5^6}}
+c_2%
{\displaystyle {\Lambda \over \kappa ^2L_5}}
, & L_5\ll 
{\displaystyle {1 \over M}}
\\ 
-c_1N_b%
{\displaystyle {3\zeta (5) \over 4\pi ^2}}
{\displaystyle {\beta \over L_5^6}}
+c_1N_f%
{\displaystyle {M^2 \over 4\pi ^2}}
{\displaystyle {\beta \over L_5^4}}
e^{-ML_5}+c_2%
{\displaystyle {\Lambda \over \kappa ^2L_5}}
, & L_5\gg 
{\displaystyle {1 \over M}}
\end{array}
\right\}  \label{e17}
\end{equation}

Again, for the case of nondegenerate fermions, one can make the replacement
given in (\ref{e15}).

Equivalently, by (\ref{e11}), the effective potential can be written in
terms of the scalar field $\varphi $,

\begin{equation}
U\sim \left\{ 
\begin{array}{ll}
C_1(N_f-N_b)%
{\displaystyle {3\zeta (5) \over 4\pi ^2}}
\beta e^{-6\sqrt{\frac 23}\kappa \varphi }+C_2%
{\displaystyle {\Lambda \over \kappa ^2}}
e^{-\sqrt{\frac 23}\kappa \varphi }, & 
\begin{array}{c}
L_5\ll \frac 1M,\text{ \thinspace or} \\ 
e^{\sqrt{\frac 23}\kappa \varphi }\ll \frac 1{2\pi RM}
\end{array}
\\ 
&  \\ 
-C_1N_b%
{\displaystyle {3\zeta (5) \over 4\pi ^2}}
\beta e^{-6\sqrt{\frac 23}\kappa \varphi } &  \\ 
+\tilde{C}_1N_f%
{\displaystyle {M^2 \over 4\pi ^2}}
\beta e^{[-M(2\pi R)\exp (\sqrt{\frac 23}\kappa \varphi )]}+C_2%
{\displaystyle {\Lambda \over \kappa ^2}}
e^{-\sqrt{\frac 23}\kappa \varphi }, & 
\begin{array}{c}
L_5\gg \frac 1M,\text{ \thinspace or} \\ 
e^{\sqrt{\frac 23}\kappa \varphi }\gg \frac 1{2\pi RM}
\end{array}
\end{array}
\right\}  \label{e18}
\end{equation}

\noindent with the replacement (\ref{e15}) for nondegenerate fermions 
\begin{equation}
N_fM^2e^{-M(2\pi R)\exp (\sqrt{\frac 23}\kappa \varphi )}\rightarrow 
\mathop{\displaystyle \sum }
_i\left\{ N_f^{(i)}M_i^2e^{-M_i(2\pi R)\exp (\sqrt{\frac 23}\kappa \varphi
)}\right\}  \label{e19}
\end{equation}

\section{``Semi-vacuumless'' Domain Walls and Bubbles}

\subsection{The Assumed Form of the Effective Potential $U$}

In what follows we assume that the potential parameters allow the low
temperature 4d effective potential to take a form (see \cite{BG}) such that $%
U$ has a local minimum at some finite value $L_{5,\min }$ (corresponding to $%
\varphi =\varphi _{\min }$), a local maximum at some finite $L_{5,\max
}>L_{5,\min }$ (corresponding to $\varphi =\varphi _{\max }>\varphi _{\min }$%
), and asymptotically, $U\rightarrow 0$ as $L_5\rightarrow \infty $ ($%
\varphi \rightarrow \infty $). Therefore, there is a vacuum state at $%
\varphi =\varphi _{\min }$ where $U=U(\varphi _{\min })$ and another low
energy state as $\varphi \rightarrow \infty $, where $U\rightarrow 0$. These
two low energy states are separated by a potential barrier at $\varphi
=\varphi _{\max }$ where $U=U(\varphi _{\max })$. We can think of this type
of system as being ``semi-vacuumless'', since it is a hybrid of a
``vacuumless'' system (see, e.g., \cite{CV,Baz}) and a system with an
ordinary vacuum state.

\subsection{The ``Semi-vacuumless'' Domain Wall}

The potential described above allows the formation of a domain wall that
interpolates between the vacuum domain ($\varphi =\varphi _{\min }$) and the
``vacuumless'' domain ($\varphi \rightarrow \infty $). We will refer to this
type of domain wall as a ``semi-vacuumless'' domain wall. This type of
domain wall differs from the ordinary $\varphi ^4$ ``kink'' type [and from a
symmetric ``vacuumless'' type (see, e.g., \cite{CV,Baz})] in that there is
no discrete symmetry associated with the potential, i.e., the potential and
the domain wall are asymmetric. Furthermore, assuming that in the
``vacuumless'' domain the scalar field assumes a finite value $\varphi _1$
with $U(\varphi _1)>0$, it is most likely that the two low energy states are
nondegenerate, with $U(\varphi _{\min })\neq U(\varphi _1)$.

Let us consider a static, planar domain wall lying in the $y-z$ plane
described by $\varphi =$ $\varphi (x)$. The 4d Einstein Frame effective
Lagrangian for $\varphi $ is ${\cal L}_\varphi =\frac 12(\partial \varphi
)^2-U(\varphi )$ and the associated energy-momentum tensor is $T_{\mu \nu
}=\partial _\mu \varphi \partial _\nu \varphi -\eta _{\mu \nu }\left[ \frac 1%
2(\partial \varphi )^2-U\right] $. This gives energy density and stress
components $T_{00}=\frac 12\varphi ^{\prime 2}+U$ , $T_{11}=\frac 12\varphi
^{\prime 2}-U$, and $T_{22}=T_{33}=-T_{00}=-\left( \frac 12\varphi ^{\prime
2}+U\right) $, where $\varphi ^{\prime }=\partial \varphi /\partial x$. In a
flat Minkowski background the equation of motion gives $\frac 12\varphi
^{\prime 2}=U+K$, where $K$ is a constant of integration. The energy density
and stress components of the domain wall configuration can then be written
as 
\begin{equation}
T_{11}=K,\,\,\,\,\,\,\,\,\,\,T_{22}=T_{33}=-T_{00}=-(2U+K)  \label{e20}
\end{equation}

\subsection{Unstable Domain Bubbles of Extra Dimensions}

With the assumption that the domain wall connects two different spatial
regions where $\varphi \approx \varphi _{\min }$ on one side of the wall and 
$\varphi \approx \varphi _1$ on the other side, with $U(\varphi _{\min
})\neq U(\varphi _1)$, (\ref{e20}) shows that the tangential stresses $%
T_{22} $ and $T_{33}$ will be different on the two different sides of the
wall, indicating that the static, planar domain wall solution is unstable
against bending. The wall will tend to bend toward the higher energy density
(i.e., higher $|T_{22}|=|T_{33}|$ ) side, and one expects a network of
bubbles to form. A bubble is surrounded by a lower energy density region and
encloses a higher energy density one. We are interested in the case where $%
\varphi _1\gg \varphi _{\min }$, that is, in the case where $\varphi _{\min
} $ takes on a small value, giving rise to a small scale factor $B_{\min
}=e^{\sqrt{\frac 23}\kappa \varphi _{\min }}$, and $\varphi _1$ takes on a
very large value so that the scale factor $B_1=e^{\sqrt{\frac 23}\kappa
\varphi _1}$ becomes large enough so that the size of the extra dimension,
characterized by the circumference $L_5=(2\pi R)B(x)$, becomes macroscopic.
In this case, the ``semi-vacuumless'' domain bubble encloses a region that
is effectively five dimensional and is surrounded by a region that is
effectively four dimensional, or vice-versa. Here, attention is focused on
the case where $U(\varphi _1)>U(\varphi _{\min })\approx 0$ (which can be
achieved by tuning the cosmological constant, for instance), so that,
effectively, the bubble encloses a 5d spacetime and is surrounded by a 4d
spacetime. Such a bubble, however, is not stable, since there is no
mechanism to balance the inward pressure caused by the bubble's surface
energy (and an assumed negligibly small volume energy), and the bubble
therefore collapses.

\section{A Stabilization Mechanism for a 4d Dimension Bubble Enclosing 5d}

\subsection{The Dimension Bubble as a Nontopological Soliton Entrapping
Bosons}

The ``dimension bubble'' described above, which is effectively surrounded by
a 4d spacetime and encloses a 5d spacetime, can be stabilized by a gas of
particles trapped within the bubble. This mechanism can arise quite
naturally if the Lagrangian ${\cal L}$ contains a complex scalar field $\chi 
$ with Lagrangian 
\begin{equation}
\begin{array}{ll}
{\cal L}_\chi & =(2\pi R){\cal L}_5,_\chi =\tilde{g}^{MN}(\partial _M\chi
)^{*}(\partial _N\chi )-V \\ 
& =e^{\sqrt{\frac 23}\kappa \varphi }g^{\mu \nu }(\partial _\mu \chi
)^{*}(\partial _\nu \chi )-V
\end{array}
\label{e21}
\end{equation}

\noindent where $V=V(|\chi |)$ and we have used, from (\ref{e4}), $\tilde{g}%
^{\mu \nu }=e^{\sqrt{\frac 23}\kappa \varphi }g^{\mu \nu }$, along with the
assumption $\partial _5\chi =0$. From (\ref{e5}), the associated 4d Einstein
Frame action is 
\begin{equation}
S_\chi =\int d^4x\sqrt{-g}e^{-\sqrt{\frac 23}\kappa \varphi }{\cal L}_\chi
=\int d^4x\sqrt{-g}{\cal L}_{\chi ,eff}  \label{e22}
\end{equation}

\noindent from which we identify an effective 4d $\chi $ Lagrangian 
\begin{equation}
{\cal L}_{\chi ,eff}=\left| \partial \chi \right| ^2-e^{-\sqrt{\frac 23}%
\kappa \varphi }V(\left| \chi \right| )  \label{e23}
\end{equation}

\noindent and an effective $\chi $ potential 
\begin{equation}
U_\chi =e^{-\sqrt{\frac 23}\kappa \varphi }V(\left| \chi \right| )
\label{e24}
\end{equation}

The $\chi $ boson mass obtained from the effective potential $U_\chi $ is 
\begin{equation}
m_\chi ^2=e^{-\sqrt{\frac 23}\kappa \varphi }\left( \frac{\partial ^2V}{%
\partial \chi ^{*}\partial \chi }\right) _{vac}=\mu ^2e^{-\sqrt{\frac 23}%
\kappa \varphi },  \label{e25}
\end{equation}

\noindent where $\mu ^2=(\partial ^2V/\partial \chi ^{*}\partial \chi
)|_{vacuum}$ is the mass parameter in the 5d theory.

For our dimension bubble enclosing a 5d spacetime (with large $\varphi $
inside the bubble) and surrounded by a 4d spacetime (with relatively small $%
\varphi $ outside the bubble), if the condition $\exp [\kappa (\varphi
_{out}-\varphi _{in})]=\exp [\kappa (\varphi _1-\varphi _{\min })]\ll 1$,
then the $\chi $ boson masses inside and outside the bubble differ
dramatically with

\begin{equation}
\frac{m_{\chi ,out}^2}{m_{\chi ,in}^2}=e^{-\sqrt{\frac 23}\kappa (\varphi
_{out}-\varphi _{in})}\gg 1  \label{e26}
\end{equation}

\noindent The result is that $\chi $ bosons get trapped inside the bubble,
as with the case of nontopological soliton (NTS) bags studied previously\cite
{NTS}. The trapped $\chi $ bosons exert an outward pressure that, at
equilibrium, can counterbalance the inward pressure of the bubble wall.

\subsection{$\chi $-Boson Stabilized Solitons}

Let us consider the case where the $\chi $ boson has a mass $m$ outside the
bag and is effectively massless inside the bag. The ground state boson
kinetic energy inside the bag can be estimated by setting the De Broglie
wavelength of a single boson equal to the diameter of the bag that confines
it: $KE\approx p=\hbar k=2\pi /\lambda \sim \pi /R$. For $Q$ bosons inside
the bag, the kinetic energy is then estimated to be $E_\chi \sim Q\pi /R$,
which agrees with the results obtained from use of a trial function by
Frieman et al. (see ref\cite{NTS}). We shall verify this result by obtaining
the solution for $\chi $ in the $\varphi $ background and calculating
explicitly the energy $E_\chi $.

For definiteness, we take the effective potential of (\ref{e24}) to be given
by $U_\chi =m_\chi ^2\chi ^{*}\chi $, where $m_\chi =m$ outside the soliton
and $m_\chi =0$ inside. For the background solution we assume $\varphi $ to
take on constant values $\varphi _{in}$ and $\varphi _{out}$ on the inside
and outside of the soliton, with a jump at the bubble wall (i.e., the thin
wall approximation). The effective Lagrangian ${\cal L}_\chi =\left|
\partial \chi \right| ^2-U_\chi $ then gives an associated energy density $%
T_{00}^\chi =\left| \partial _0\chi \right| ^2+\left| \partial _r\chi
\right| ^2+U_\chi $ for a spherically symmetric soliton. The energy $%
E=E_\chi +E_{wall}$ of the soliton is then $E=4\pi \int T_{00}^\chi
r^2dr+4\pi R^2\Sigma $, where $R$ is the soliton radius and $\Sigma $ is the
surface energy density of the bubble wall. To evaluate $E$, we first need to
obtain the solution for $\chi $.

We take $\chi (r,t)=F(r)\,e^{i\omega t}$. The field equation for $\chi $
gives 
\begin{equation}
\frac{d^2F}{dr^2}+\frac 2r\frac{dF}{dr}+(\omega ^2-m_\chi ^2)F=0  \label{a1}
\end{equation}

\noindent and the conserved charge of $Q$ bosons trapped inside the soliton
is 
\begin{equation}
Q=8\pi \omega \int_0^\infty F^2r^2dr  \label{a2}
\end{equation}

\noindent For a finite energy solution, we require $F$ to vanish outside the
bubble for the case that $(\omega ^2-m^2)\leq 0$, while for $(\omega
^2-m^2)>0$ we have 
\begin{equation}
F\propto \frac{e^{-(m^2-\omega ^2)^{1/2}r}}{(m^2-\omega ^2)^{1/2}r}
\label{a3}
\end{equation}
\noindent so that $F$ rapidly approaches zero outside the soliton. We
therefore take $F=0$ as an approximate solution for $r>R$. The regular
solution for $F$ inside the soliton is $F=F_0\sin (\omega r)/\omega r$. The
approximate solution is therefore 
\begin{equation}
F=\left\{ 
\begin{array}{cc}
F_0\frac{\sin (\omega r)}{\omega r}, & (r<R) \\ 
0, & (r>R)
\end{array}
\right\}  \label{a4}
\end{equation}

\noindent For continuity of the solution at the bubble wall $r=R$, we set 
\begin{equation}
\omega R=\pi  \label{a5}
\end{equation}

The solution (\ref{a4}) and the boundary condition (\ref{a5}) give the
energy $E_\chi =4\pi ^2F_0^2/\omega $ and boson charge $Q=4\pi
^2F_0^2/\omega ^2$, or 
\begin{equation}
E_\chi =\omega R=\frac{Q\pi }R  \label{a6}
\end{equation}

\noindent as obtained in the quick estimate above.

We have assumed the vacuum energy density to vanish inside the soliton, so
that (neglecting gravitational effects) the energy of the NTS is determined
by the contributions from the domain wall surface energy and the energy of
the entrapped bosons,

\begin{equation}
E=\frac{Q\pi }R+4\pi \Sigma R^2  \label{e27}
\end{equation}

\noindent where $\Sigma $ is the surface energy density of the wall.
Minimization of this energy gives the NTS equilibrium radius 
\begin{equation}
R=\frac 12\left( \frac Q\Sigma \right) ^{1/3}  \label{e28}
\end{equation}

\noindent From (\ref{e27}) and (\ref{e28}) the equilibrium energy, or mass,
of the bubble is 
\begin{equation}
E=3\pi (Q^2\Sigma )^{1/3}  \label{e29}
\end{equation}

We can obtain approximate stability conditions for the bosonic charge $Q$
and the bubble radius $R$ by requiring the $\chi $ bosons to remain trapped
inside the bubble rather than escaping to the outside. For $Q$ free bosons
of mass $m$ outside the bag, the minimal energy is $E_{Q\,free}=Qm$.
Therefore, for the NTS bubble stability we require the equilibrium bubble
energy $E<E_{Q\,free}$ which, by (\ref{e29}), implies that

\begin{equation}
Q^{1/3}>3\pi \frac{\Sigma ^{1/3}}m  \label{e30}
\end{equation}

\noindent By (\ref{e28}) this translates into the condition for the bubble
radius that 
\begin{equation}
R>\frac{3\pi }{2m}  \label{e31}
\end{equation}

\noindent for a stable bubble.

\smallskip 

\section{Summary}

We have considered a type of case similar to that previously studied by Blau
and Guendelman\cite{BG} wherein a scalar field (appearing within a
dimensionally reduced 4d effective theory)--that is associated with the
scale factor of an extra (compact) dimension in a 5d theory--forms a domain
wall interpolating between two different spatial regions. When the scalar
field takes on radically different values in the two different domains, the
effective dimensionality of the space also becomes different in these
different domains. [As an example, we have followed\cite{BG} in constructing
a form of 4d effective potential arising from the low temperature Rubin-Roth
potential\cite{RR} (due to one loop quantum contributions from massless
bosons and massive or massless fermions) along with a cosmological
constant.] Domain walls which are unstable against bending can give rise to
a network of closed ``dimension bubbles'', where the effective spacetime
dimensionality is different in the interior and exterior regions of a
bubble. Without a stabilization mechanism, these bubbles collapse due to the
tension in the bubble walls.

However, the inclusion of a scalar field $\chi $ in the original 5d theory
can give rise to a stabilization mechanism in a rather natural way. In
essence, due to the different values of the conformal factor in the interior
and exterior regions of the bubble, the $\chi $ boson mass can change
drastically across the bubble wall. We have considered the case of a bubble
surrounded by a 4d region and enclosing a 5d region, with $\chi $ bosons
that are massive outside the bubble, but are effectively massless inside the
bubble. Therefore, the $\chi $ particles are effectively trapped inside the
bubble, exerting an outward pressure which can stabilize the bubble from
total collapse, provided that certain model parameters lie within
appropriate ranges. This resembles the stabilization mechanism for
``ordinary'' 4d nontopological solitons (NTSs) studied previously by
Frieman, Gleiser, Gelmini and Kolb\cite{NTS}.

Semi-vacuumless domain walls and dimension bubbles can also arise from the
4d effective potential associated with the model of a {\it classical}
stabilization of {\it two} extra dimensions in a spherically reduced 6d
model described by Carroll, Geddes, Hoffman and Wald\cite{CGHW} (see sec. 3
of that paper). In that model, an extra dimensional magnetic field, along
with a cosmological constant and the curvature term of the extra dimensional
two-sphere, give rise to a 4d effective potential [given by eq.(36) of ref.%
\cite{CGHW}] with the same basic shape as the effective potential $U$
arising from quantum fluctuations considered here [see eq. (\ref{e9})
above]. These bubbles could also be stabilized by the same mechanism of $%
\chi $ boson entrapment.

As in the case of the ordinary NTSs, dimension bubbles, if they exist, could
contribute to the dark matter of our (mostly) 4d universe.

\smallskip 

\newpage\

\end{document}